\documentclass[aps,twocolumn,showpacs,preprintnumbers,amsmath,amssymb,floatfix]{revtex4}
\usepackage{graphicx}% Include figure files
\usepackage{dcolumn}% Align table columns on decimal point
\usepackage{bm}% bold math
\usepackage[german,american]{babel}

\newcommand{\figref}[1]{Fig.\ \ref{#1}}
\newcommand{\dwp}{DWP}
\newcommand{\stm}{STM}

\begin{document}

%\preprint{APS/123-QED}

\title{What is moving in silica at 1 K? A computer study of the low-temperature anomalies}

\author{J. Reinisch}
 %\\altaffiliation[Also at ]{Physics Department, XYZ University.}%Lines break automatically or can be forced with \\
\author{A. Heuer}%
 %\\email{Second.Author@institution.edu}

\affiliation{
Westf\"{a}lische Wilhelms-Universit\"{a}t M\"{u}nster, Institut f\"{u}r Physikalische Chemie
Corrensstr. 30, 48149 M\"{u}nster, Germany }

\date{\today}% It is always \today, today,

\begin{abstract}
Though the existence of two-level systems (TLS) is widely accepted
to explain low temperature anomalies in many physical observables,
knowledge about their properties is very rare. For silica which is
one of the prototype glass-forming systems we elucidate the
properties of the TLS via computer simulations by applying a
systematic search algorithm. We get specific information in the
configuration space, i.e. about relevant energy scales, the
absolute number of TLS and electric dipole moments. Furthermore
important insight about the real-space realization of the TLS can
be obtained. Comparison with experimental observations is
included.
\end{abstract}

 %PACS, the Physics and Astronomy
\maketitle

Most kinds of disordered solids show anomalous behavior at very
low temperatures (Kelvin regime and below) as compared to their
crystalline counterparts. Many of the observed features can be
explained by the Standard Tunneling Model (\stm)
\cite{Phillips:1972,Anderson:1972} and its generalization, which is
the Soft-Potential Model \cite{Karpov:1983,Buchenau:1991}. The
basic idea of the \stm\ is to postulate the possibility of
localized transitions between different configurations, i.e.
adjacent minima of the potential energy landscape. Such a
transition can be described by a double-well potential (DWP),
characterized by an asymmetry $\Delta$, potential height V and
distance $d$ between both configurations. From a
quantum-mechanically perspective at low temperatures the system is
tunneling between both configurations and the DWP is characterized
by the lowest two eigenstates. If their energy difference E is in
the Kelvin regime, these DWP may contribute to the low-temperature
anomalies. Then one may speak of Two Level Systems (TLS). The
TLS can couple to strain and electric fields and therefore show up
in observables like thermal conductivity, sound absorption and
dielectric response \cite{Esquinazi:1998}. Recently, even
observations about the interaction of different TLS have been
reported \cite{Natelson:1998,Classen:2000,Rosenberg:2003}.

So far it has not been possible to derive a theory of the
low-temperature anomalies of real glass-forming systems from first
principles, except for mean-field models \cite{Kuehn:2003}
and random first order transition theory \cite{Lubchenko:2001}.
Thus for a prototype system like SiO$_2$ the STM has to be
basically considered as phenomenological. Important questions
emerge. (Q1) How many TLS are present? This is, of course, a
central experimental observables. (Q2) What are the energetic
properties of the TLS, e.g. typical barrier heights? (Q3) What is
the {\it average} microscopic nature of the TLS?

Computer simulations may help to shed some light on the nature of
the low-temperature anomalies. In previous work on SiO$_2$ the
trajectories, generated either by molecular dynamics
\cite{Trachenko:1998, Trachenko:2000} or by the
activation-relaxation technique \cite{Mousseau:2000}, have been
analyzed with respect to transition events between different
structures. Both approaches yield some interesting insight into
the nature of relaxation processes in SiO$_2$. (Q1), however,
requires a systematic search procedure, and (Q2) and (Q3) a
sufficiently large number of characteristic DWP and thus an
efficient search method.  This has not been the scope of previous
simulations on SiO$_2$.

In recent years we have developed a set of simulation techniques
which allow us to approach these questions
\cite{Reinisch_jltp:2004,Reinisch_prb:2004}.  (Q1) Starting from
representative low-energy structures we {\it systematically}
search for adjacent minima of the potential energy landscape, i.e.
DWP. It is possible to formulate an intrinsic completeness
criterion to check whether an existing adjacent minimum has indeed
been found \cite{Reinisch_jltp:2004}. If this criterion is
fulfilled we have access to the {\it absolute} number of TLS.
Saddles are determined via a robust saddle-search algorithm
\cite{Doliwa:2003a}. (Q2) The typical interatomic energy scale and
thus the typical range of DWP asymmetries is of the order 1 eV.
Thus via a direct search it is impossible to find a set of DWP
with asymmetries in the Kelvin regime. In the Soft-Potential Model
a DWP is parametrized by a quartic polynomial $\sum_{n=2}^4 w_n
x^n$, thereby reproducing the values for $\Delta,V$ and $d$. If
not mentioned otherwise, $d$ corresponds to the mass-weighted
distance, obtained from a straight connection of both minima with
the transition state. Using this parametrization we can first
determine distribution functions $p_i(w_i)$ from our set of
numerically found DWP and then generate an arbitrary large number
of DWP with the same statistics.  Thus it is possible to estimate
the DWP distribution function $p(d,V,\Delta)$ over a broad range
of parameters and in particular to get important information about
the properties of nearly symmetric DWP, i.e. TLS. (Q3) Due to the
effectiveness of the search algorithm we find a sufficiently large
number of DWP to perform a reasonable statistical analysis with
respect to microscopic properties. All technical aspects and in
particular the justification of the parametrization approach can
be found in \cite{Reinisch_jltp:2004,Reinisch_prb:2004}. So far we
have successfully applied these methods to Lennard-Jones model
systems \cite{HeuerSilbey:1996, HeuerSilbey:1993a,
Doliwa:2003a,Reinisch_jltp:2004,Reinisch_prb:2004}; see also
\cite{Oligschleger:1995} and chapter 8 of Ref.\cite{Esquinazi:1998}.

In this paper these techniques are for the first time applied to
pure SiO$_2$, modelled by the BKS-potential \cite{BKS:1990}.
Despite the enormous numerical effort, involved in these
calculations, we were able to obtain detailed results about the
central questions (Q1)-(Q3), introduced above.

\begin{figure}
  \includegraphics[width=4.25cm]{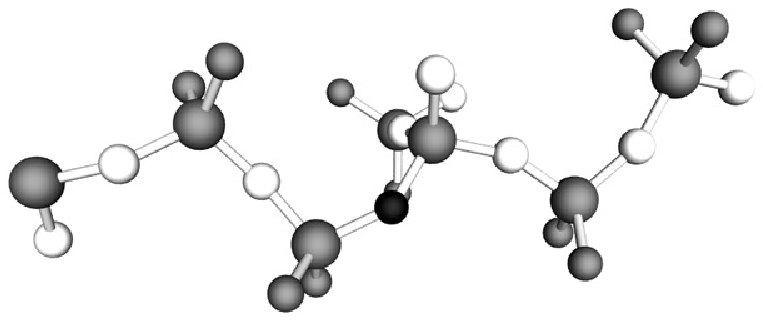}
  \includegraphics[width=4.25cm]{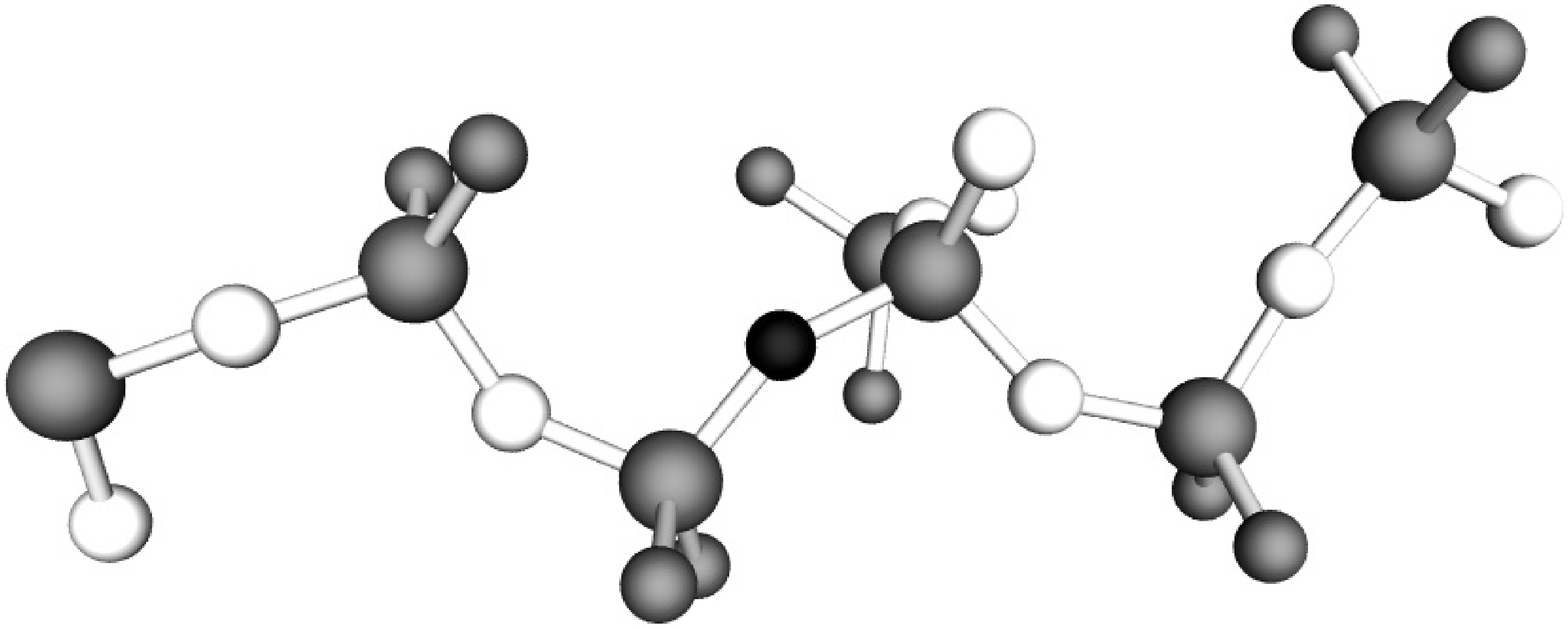}
  \caption{\label{fig_example}
  The two structures correspond to the minima of a \dwp . The black particle
  is the one, moving most, and the white particles mark the 11 most displaced ones, of which
  two oxygens are not shown as they are not connected to the above network.
  The grey particles correspond to the less moving oxygen (small spheres)
  and silicon (large spheres) atoms.}
\end{figure}

We have analyzed system sizes of 150 and 600 particles with the
standard density of 2.3 $\mathrm{g/cm^3}$ and a short-range cutoff
for the BKS-potential of 8.5 \AA. The starting configurations for
the systematic search correspond to equilibrium configurations at
3000 K, which subsequently were minimized. We have obtained
starting structures with and without defects, i.e. deviations from
a perfect tetrahedral coordination. It turned out that the
properties of DWP are totally different in both cases because a
defect is very likely related to DWP with high barriers
\cite{Mousseau:2000} and a stronger spatial localization.
Experimental silica samples, however, freezing in at the
calorimetric $T_g$, do not show a high number of defects
\cite{Horbach:1999}. Therefore we have only considered structures
without defects. Actually, these non-defect structures are very
similar to those one would obtain from simulations at much lower
temperatures \cite{aimorn}. Furthermore we have only taken into
account DWP with $|\Delta| < 1500$ K. This already corresponds to
the symmetric side of the asymmetry distribution
\cite{Mousseau:2000}. With this constraint we have been able to
produce a set of 250 DWP for 150 particles and 50 DWP for 600
particles. One example is shown in \figref{fig_example}. On
average we have found one DWP per 14 defect-free starting
structures. The completeness criterion is fulfilled. To a good
approximation the distribution of asymmetries is constant (data
not shown). Thus the number of DWPs with $|\Delta| < 1$ K can be
estimated via $1/(14 \cdot k_B 1500 \mathrm{\ K} \cdot L^3) \approx 5 \cdot
10^{44}/(\mathrm{Jm^3})$ where $L$ is the length of the simulation box.

For the 600 particle system the search procedure is complicated by
the occurrence of independent DWP, which makes a systematic search
much more time-consuming. Due to the better statistics  we report
results for the 150 particle system.  We checked, however, that
apart from minor variations in the participation ratio (see below)
all properties, discussed in this work, are identical for both
system sizes within statistical uncertainty. This is compatible
with our previous result that the thermodynamics of BKS-SiO$_2$
with only 99 particles is, apart from trivial scaling, basically
identical to that of a macroscopic system \cite{aimorn}.

\begin{figure}
  \includegraphics[width=8.6cm]{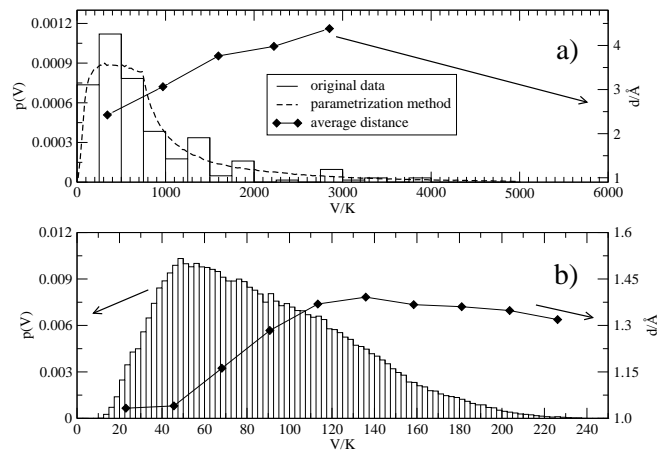}
  \caption{\label{fig_barrier}
 a) Distribution of barrier heights for the simulated data with
   asymmetries up to $1500 \mathrm{\ K}$. Furthermore the distribution, generated
    from our statistical analysis, is shown (see text).
   b) The calculated distribution of barrier heights for TLS.
   In both cases we also show the average distance between two minima for a given
   barrier height.
  }
\end{figure}

In Fig.~\ref{fig_barrier}a we show the distribution of barrier
heights. Using the parametrization method we obtain the full
distribution of DWP, i.e. $p(d,V,\Delta)$. Interestingly, the
typical barrier heights agrees well with the value of approx. 500 K,
estimated from Brillouin scattering experiments
\cite{Tielbuerger:1992}. Using the same constraint $|\Delta| <
1500$ K the original distribution is reproduced; see also
Fig.~\ref{fig_barrier}a.  Next we have estimated the tunneling
matrix element for every DWP in the full distribution
$p(d,V,\Delta)$, using the WKB approximation \cite{Phillips:1987},
and calculated the energy splitting $E$ of the two lowest
eigenfunctions and the relaxation rate $\tau$
\cite{Phillips:1987}. The subset of DWP with $E < 2$ K and $\tau <
5$s, which from now on we denote TLS, is relevant in the Kelvin
regime. The distribution of barrier heights of the TLS is
presented in Fig.~\ref{fig_barrier}b. This distribution shows a
 maximum around 50 K. The average distance between both
minima varies with increasing potential height. More generally,
one observes that very nearby minima have smaller asymmetries and
barrier heights. This correlation has been also observed for the
Lennard-Jones system \cite{Reinisch_jltp:2004,Reinisch_prb:2004}.

The effective density of states $P_{\mathrm{eff}}$ is accessible,
i.e, by sound absorption experiments \cite{Berret:1988}. Based on
our distribution $p(d,V,\Delta)$ and our systematic search
procedure we can estimate this value from first principles. Again,
for the necessary technicalities we have to refer the reader to
the corresponding analysis of Lennard-Jones systems
\cite{Reinisch_jltp:2004,HeuerSilbey:1993a}.  We obtain for pure
defect free silica that
\begin{equation}
 P_{\mathrm{eff}}(E)=1.3 \cdot 10^{44} (E/1\mathrm{\ K})^{0.08}/\mathrm{(J m^3)}
\end{equation}
which is of the same order as the value from the simple
estimation, reported above.  $P_{\mathrm{eff}}(E\approx 1$ K) can be
directly compared to experimental \cite{Sahling:2002} values
between $(4 - 7)\cdot 10^{44}/\mathrm{(Jm^3)}$ for Suprasil W (205 ppm
impurities) and $8\cdot 10^{44}/\mathrm{(J m^3)}$ for Suprasil I(1250 ppm
impurities). The dependence on energy is weak but significant and
can be regarded as a correction to the \stm. Although for a first
principle analysis the agreement of $P_{\mathrm{eff}}(E=1$ K) with the
experimental data is already remarkable the remaining differences
can be (at least partly) related to different aspects: (i) Using a
Frank-Condon factor for the multidimensional renormalization and
taking into account the entropic effects, obtained from a harmonic
analysis of the transition state and the minima, may increase the
presented value by a factor of 1.5 according to our estimation.
(ii) The remaining defects,
present even at $T_g$, may have some impact on this number. Since
for pure SiO$_2$ the number of defects is only roughly known it is
difficult to estimate this contribution \cite{Horbach:1999}.
Since, however, defects are very efficient in forming DWP (data
not shown) even a concentration of 100 ppm contributes to
$P_{\mathrm{eff}}$. (iii) The impurities in real silica systems contribute
additional extrinsic DWP.

In the remaining part of this paper we would like to discuss the
real-space properties of the DWP found in our simulation. For the
DWP in \figref{fig_example} the tetrahedra, involved in the main
displacement, basically form a 1D-chain. Visual inspection of
about 50 DWP indicates that this reduced dimensionality of the
transition is a generic feature. One can also see that the number
of participating particles is rather small. Interestingly, only
two oxygens per tetrahedron move significantly. In what follows we
will present a statistical analysis of all DWP to extract the
average behavior.

For the average participation ratio different definitions can be
used \cite{Reinisch_jltp:2004}. We obtain (both for N=150 and
N=600) $\langle d^2/d^2_{max}\rangle \approx 1/\langle
d^2_{max}/d^2\rangle \approx 9$, $\langle d^4/\sum_i d^4_i \rangle
\approx 24 (N=150)$ and $\langle d^4/\sum_i d^4_i \rangle \approx
31 (N=600)$. For this specific analysis $d$ describes the
euclidian distance between two configurations.  $d_i$ is the
displacement of particle $i$ and $d_{max}$ the displacement of the
most displaced particle. Actually, very similar values are
reported in \cite{Trachenko:2000} and in \cite{Ekunwe:2002} where
the tetrahedra relaxation upon applied pressure has been
monitored. The most displaced particle is oxygen in all analyzed
cases (which is further denoted as central oxygen) and the 4 most
displaced particles are exclusively oxygen in over $99\%$ of all
analyzed cases. Averaged over all particles, the mean square
displacement of oxygen is 2.3 times larger than that of silicon.
Furthermore, we have checked that the participation ratio
decreases with decreasing distance in analogy to the Lennard-Jones
system \cite{HeuerSilbey:1993a, Reinisch_jltp:2004}. One can
estimate that the above values for the participation ratio
decrease by approx. 25\% when evaluated for distances of $d \approx
1.2$ \AA \, for typical TLS rather than $d \approx 2.8$ \AA \, for
typical DWP (see Fig.~\ref{fig_barrier}). Furthermore, the TLS
have an electric dipole moment which is also accessible
experimentally. Using the partial charges of the BKS potential we
obtain an average dipole moment of 0.65 D which is in excellent
agreement with the experimental value of 0.5 D
\cite{Golding:1979}.

In the next step we have analyzed the average local environment of
the central oxygen. It turns out that before and after the flip
its average Si-O-Si angle is $(148.5 \pm 1)^\circ$ whereas at the
transition state it is approx. 160$^\circ$. Thus to a good
approximation the Si-O-Si bond performs a flip rather than a
simple rotation. Actually, the average Si-O-Si angle for all bonds
is 152$^\circ$ in the BKS-system \cite{Yuam:2003}, which indicates
a structural anomaly around the central oxygen. The observed
strong dependence of the density of states on the
inter-tetrahedral structure \cite{Liu:1995} fits in this picture,
since the relevant Si-O-Si angle is the most important parameter
for the inter-tetrahedral structure.

Furthermore, also the O-Si-O angles vary during the transition.
The most displaced tetrahedra shows variations of the O-Si-O
angles of  2.7$^\circ$ which is nearly half of the overall
equilibrium variation of intra-tetrahedral angles; see also
\cite{Ekunwe:2002}. However, for this quantity no structural
anomalies could be observed. One may speculate that these
distortions are an important process of dampening the motion and
thus to localize the DWP.
\begin{figure}
  \includegraphics[width=8.6cm]{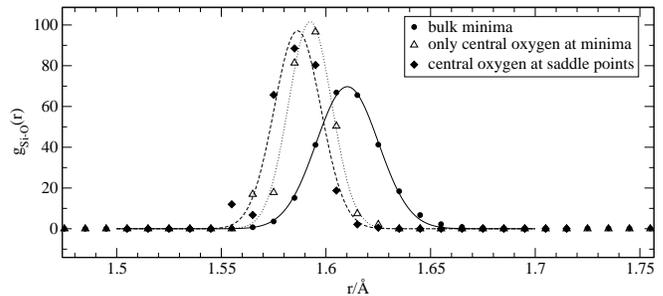}
  \caption{\label{fig_gofr}
           Special radial distribution function for the central particles of the \dwp\ in
           comparison to a general $g(r)$. No
           differences can be seen beyond the first peak.}
\end{figure}
Another structural anomaly can be seen from  comparing the radial
distribution functions around the central oxygen at the transition
and two minima positions with that of an average oxygen; see
\figref{fig_gofr}. First, we observe a shortening of the Si-O bond
during the transition. Thus the Si-O-Si bond flip is achieved by a
squeezing of the oxygen through the two adjacent Si atoms. Second,
and even more important, the Si-O bond lengths for the typical DWP
minima is significantly smaller than observed for the bulk, and
their distribution is narrower.

In order to minimize the energy variation during the transition it
is essential that the total amount of particle displacement is as
small as possible. Thus tetrahedra rotations around the threefold
axis, giving three equally large oxygen displacements, should be
very unfavorable. A better choice would be rotation around one of
the tetrahedral edges, thus moving only two oxygen atoms. This
type of  motion has already been indicated in
\figref{fig_example}, where only two oxygens at most tetrahedra
are largely displaced.

\begin{figure}
  \includegraphics[width=8.6cm]{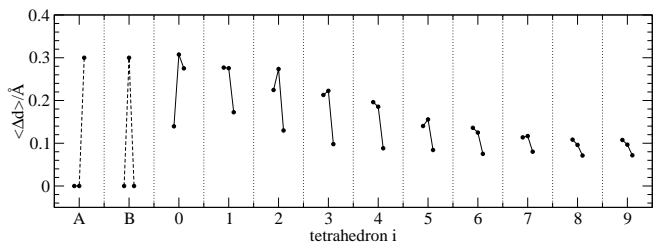}
  \caption{\label{fig_scheme}Differences of the displacements of the four oxygens at the 10 most
           displaced tetrahedra averaged over all DWP.
            Within each tetrahedron the
           left point displays the displacement difference ($\Delta d_{0,1}$)
           between the most displaced
           oxygen and the second most displaced oxygen. The middle point represents $\Delta d_{1,2}$
           and the right point $\Delta d_{2,3}$.
           The leftmost (A) sequence of points corresponds to a threefold rotation.
           The B sequence corresponds to a rotation around a tetrahedra edge.}
\end{figure}

To establish a statistical picture of the typical tetrahedron
rotation the average displacements of the 4 oxygen atoms at each
tetrahedron have been investigated. In \figref{fig_scheme} results
are presented for the 10 most displaced tetrahedra averaged over
all TLS. To give a better impression of the rotation the three
differences of the displacements are shown. The first dot in each
section corresponds to the difference of the displacements of the
two most displaced oxygens and so on.
In agreement with expectation the nature of the rotations is quite
different to case A (rotation around threefold axis).
Interestingly, the most displaced tetrahedron indeed shows a
rotation with some similarity to the proposed rotation axis along
a tetrahedral edge (case B in \figref{fig_scheme}). All other
rotations follow a different scheme, which  involves four
different displacements for the oxygen atoms. This can be possibly
interpreted as a transition to a statistical (low-amplitude)
displacement with no symmetry criterion for the rotation axis.

One can compare the observed rotation scheme with the proposed
motion of five connected tetrahedra by Buchenau et al. which is
often used as a microscopic visualization for SiO$_2$-TLS
\cite{Buchenau:1984}. The different displacements for all oxygens
connected to a tetrahedron as well as the Si-O-Si flip motion
fully agree with our results. There are, however, some important
modifications suggested by this work: (i) the individual
tetrahedron rotation axes does not always go through an oxygen or
silicon atom, (ii) tetrahedron distortion should be considered,
(iii) the spatial structures are one dimensional rather than three
dimensional. It might be also interesting to compare these modes
in detail with the localized vibrational modes studied, e.g., in
\cite{Taraskin:1999}.

In summary, for the first time it is possible to obtain a
systematic numerical description of the TLS for BKS-SiO$_2$. The
density of TLS and the electric dipole moment are in
semi-quantitative agreement with experimental data. Important
microscopic properties of the TLS like structural anomalies and
the nature of the displacements have been gained. In principle,
microscopic information about TLS is also available from recently
performed experiments on magnetic field dependent polarization
echoes. Thus, with the complementary views from experiment and
simulation a detailed microscopic picture of the low-temperature
anomalies will hopefully emerge.

We like to thank U. Buchenau, C. Enss, B. Doliwa, H. Lammert, A.
Saksaengwijit, H.R. Schober and M.Vogel for fruitful discussions
and the NRW International Graduate School of Chemistry for
funding.

\bibliography{reference}% Produces the bibliography via BibTeX.

\end{document}